\documentclass[aps,super,twocolumn,nofootinbib,amssymb,amsmath,footinbib]{revtex4}
\usepackage{graphicx}
\usepackage{dcolumn}
\usepackage{bm}
\usepackage[sort&compress]{natbib}
\usepackage{textcomp}
\usepackage{gensymb}

\newcommand{\Hp}{$\mathrm{H^+}$}

\begin{document}

\title{Induction kinetics of a conditional pH stress response system in {\it Escherichia coli}}
\author{Georg Fritz$^1$}
\author{Christiane Koller$^2$}
\author{Korinna Burdack$^2$}
\author{Larissa Tetsch$^2$}
\author{Ina Haneburger$^2$}
\author{Kirsten Jung$^2$} 
\author{Ulrich Gerland$\,^{1,*}$}
\affiliation{$^1$ Arnold-Sommerfeld Center for Theoretical Physics (ASC) and Center for NanoScience (CeNS), LMU M\"unchen, Theresienstr. 37, 80333 M\"unchen, Germany}
\affiliation{$^2$ Munich Center of integrated Protein Science CiPSM at the Department of Biology I, Microbiology, LMU M\"unchen, Gro§haderner Stra\ss e 2-4, 82152 Planegg-Martinsried, Germany}
\affiliation{$^*${\small {\it Corresponding author, Email address:} {\tt ulrich.gerland@physik.lmu.de}}}

\begin{abstract}
The analysis of stress response systems in microorganisms can reveal molecular strategies for regulatory control and adaptation. Here, we focus on the Cad module, a subsystem of {\it E. coli}'s response to acidic stress, which is conditionally activated at low pH only when lysine is available. When expressed, the Cad system counteracts the elevated \Hp concentration by converting lysine to cadaverine under the consumption of \Hp, and exporting cadaverine in exchange for external lysine. Surprisingly, the {\it cad} operon displays a transient response, even when the conditions for its induction persist. To quantitatively characterize the regulation of the Cad module, we have experimentally recorded and theoretically modeled the dynamics of important system variables. We establish a quantitative model that adequately describes and predicts the transient expression behavior for various initial conditions. Our quantitative analysis of the Cad system supports a negative feedback by external cadaverine as the origin of the transient response. Furthermore, the analysis puts causal constraints on the precise mechanism of signal transduction via the regulatory protein CadC. \\

\end{abstract}

\maketitle

\section*{Introduction}

During their natural life cycle, gastrointestinal bacteria are faced with acid stress while passing the extreme low pH of the stomach and being exposed to volatile fatty acids in the intestine. {\it Escherichia coli}'s remarkable ability to sustain growth over multiple decades of \Hp concentrations  \cite{Gale_BiochemJ_42} and its potential to survive extremely low pH is implemented by a battery of pH homeostasis \cite{Booth_MicrobiolRev_85, Park_MolMicrobiol_96, Richard_JBacteriol_04} and acid tolerance systems \cite{Castanie-Cornet_JBacteriol_99, Booth_AntonieVanLeewenhoek_02, Merrell_CurrOpinMicrobiol_02, Foster_NatRevMicrobiol_04}. In recent years, it was increasingly recognized that each of these subsystems is specifically activated under certain environmental conditions \cite{Booth_AntonieVanLeewenhoek_02, Foster_NatRevMicrobiol_04}, while the orchestration of the different responses is just beginning to be explored. However, a system-level study of the acid stress response requires a detailed quantitative analysis of the individual modules. 

\begin{figure}
\centerline{\includegraphics[width=8cm]{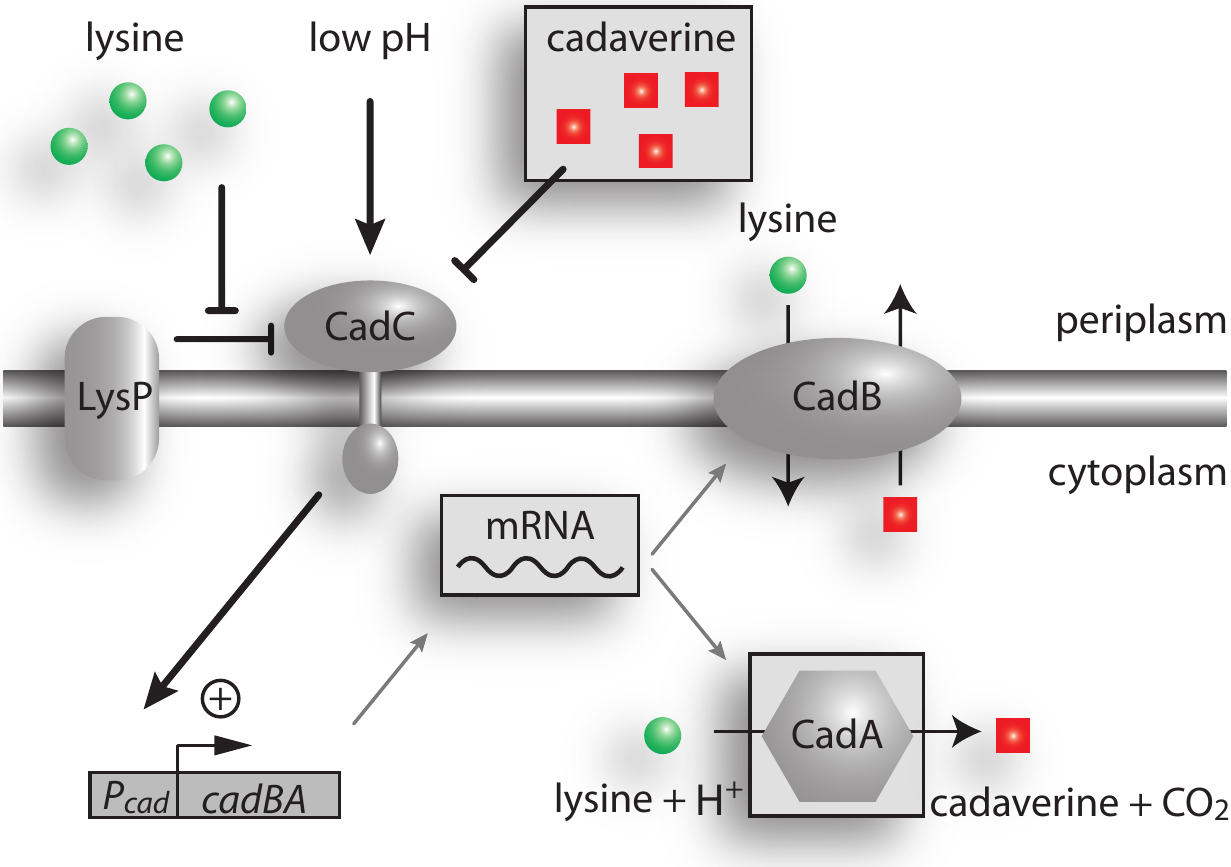}}
\caption{\label{FIGscheme} Qualitative Model of the Cad System in {\it E. coli} (simplified). The Cad system is conditionally activated by low pH and high lysine levels. Lysine inhibits the repressive effects of the lysine permease LysP on the receptor CadC, while low pH activates CadC directly. The active form of CadC activates transcription of {\it cadBA}, encoding the lysine decarboxylase CadA and the lysine/cadaverine antiporter CadB. The Cad system imports lysine, decarboxylates it under consumption of a cytoplasmic proton and exports the product cadaverine in exchange for another lysine molecule. The net effect of these reactions is the expulsion of a proton from the cytoplasm. Finally, it is believed that external cadaverine deactivates CadC. In our experiments we recorded the time-evolution of the variables in grey boxes.}
\end{figure}

One of the conditional stress response modules is the Cad system \cite{Popkin_JBacteriol_80, Meng_JBacteriol_92a, Meng_JBacteriol_92b, Neely_JBacteriol_94, Neely_JBacteriol_96}, which is induced only when acidic stress occurs in a lysine-rich environment. The three principal components of the Cad system are the enzyme CadA, the transport protein CadB, and the regulatory protein CadC, see Fig.~\ref{FIGscheme}. The decarboxylase CadA converts the amino acid lysine into cadaverine, a reaction which effectively consumes {\Hp } \cite{Sabo_Biochemistry_74}. The antiporter CadB imports the substrate, lysine, and exports the product, cadaverine. Together, CadA and CadB reduce the intracellular {\Hp } concentration and thereby contribute to pH homeostasis \cite{Meng_JBacteriol_92a, Soksawatmaekhin_MolMicrobiol_04}. The cytoplasmic membrane protein CadC not only senses the external conditions  \cite{Watson_JBacteriol_92, Neely_JBacteriol_94, Dell_MolMicrobiol_94}, but also regulates the response by binding directly to the DNA and activating the transcription of {\it cadBA} \cite{Kueper_JMolMicrobiolBiotechnol_05}. Like other members of the ToxR family \cite{Pfau_JBacteriol_98}, CadC thereby performs signal transduction in a single component, without the phosphorylation step employed by two-component systems \cite{Laub_AnnuRevGenet_07}. Fig.\,\ref{FIGscheme} also depicts the lysine permease LysP, which is not part of the {\it cad} operon, but essential for its function, since CadC senses lysine indirectly via interaction with LysP \cite{Watson_JBacteriol_92, Neely_JBacteriol_94, Dell_MolMicrobiol_94, Tetsch_MolMicrobiol_08}. In contrast, the external (periplasmic) pH is believed to be sensed directly by CadC, through a pH-dependent conformational transition \cite{Dell_MolMicrobiol_94, Lee_JBacteriol_08}. The signal integration performed by CadC then assures that CadA and CadB are produced only under the appropriate external conditions of low pH and lysine abundance. 

However, CadC also senses a third input, which seems surprising from a physiological point of view: External cadaverine binds to CadC \cite{Tetsch_MolMicrobiol_08} and represses the long-term expression of the {\it cad} operon \cite{Neely_JBacteriol_94}. As cadaverine is the end product of the decarboxylase reaction, it was suggested that it accumulates in the medium and causes a delayed transcriptional down-regulation of {\it cadBA} expression \cite{Neely_JBacteriol_96}. Although many stress response systems display a similar transient response \cite{Jung_JMolMicrobiolBiotechnol_02, Klipp_NatBiotechnol_05}, their regulation strategy appears to be fundamentally different: For instance, the osmo-stress response of yeast directly follows its stimulus (low osmolarity) and remains active until the osmolarity returns back to physiological levels.

In this study, we explore whether a negative feedback via external cadaverine can account for the transient response of the Cad system on a quantitative level. In particular, we are interested in the factors that determine the {\em duration} and the {\em amplitude} of the transient response, and ask how the addition of external cadaverine affects these characteristic quantities. It is known for instance, that external cadaverine reduces the long-term activity of the Cad system \cite{Neely_JBacteriol_94}, but one would like to know whether it shortens the duration of the transient expression pulse or reduces the amplitude of the pulse. Or does it affect both of these properties? 

To address these questions, we quantitatively measured the dynamics of the Cad system in three important variables at high time resolution: the {\it cadBA} transcript, the activity of the lysine decarboxylase CadA, and the concentration of excreted cadaverine.  Based on the existing qualitative model, cf. Fig.~\ref{FIGscheme}, we formulate a quantitative model for the Cad system, and test its agreement with the experimental response dynamics. We find that our quantitative model coherently describes the dynamical response of the wild-type Cad system within a physiological parameter regime. The available data constrains the key biochemical parameters to a narrow regime. For instance, we infer the effective {\it in vivo} deactivation threshold for the Cad system and compare it to a previously measured {\it in vitro} binding threshold \cite{Tetsch_MolMicrobiol_08}. Using the quantitative model, we formulate predictions for the response dynamics of the Cad system under conditions with initially added cadaverine, and in a mutant strain with a defective lysine permease LysP. The successful experimental validation of these predictions strongly supports the existence of the postulated feedback inhibition mechanism via cadaverine in the Cad system. Finally, we discuss the causal constraints of our results on the signal transduction mechanism by CadC, helping to discriminate between two contradicting models.

\section*{Results}

\subsection*{Transient expression dynamics}

\begin{figure}[t]
\centerline{\includegraphics[width=6.5cm]{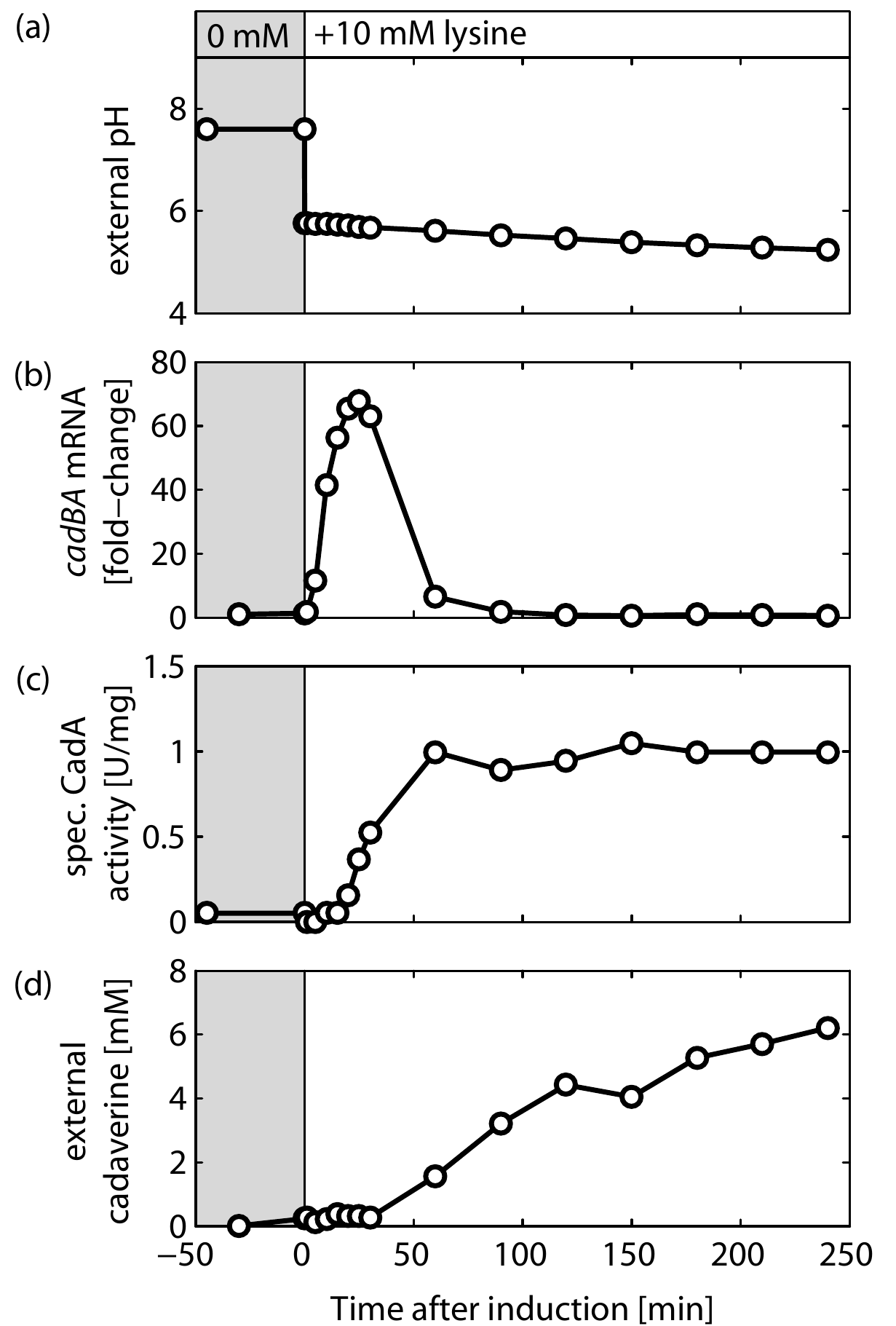}}
\caption{\label{FIGtimeseries_exp} 
Induction kinetics of the {\it cad} operon in {\it E. coli} MG1655. (a) The Cad system was induced at t=0\,min by a shift from pH 7.6 to pH 5.8 and simultaneous addition of 10 mM lysine. The time evolution of the {\it cadBA} mRNA (b), the CadA activity (c) and the extracellular cadaverine concentration (d) was determined as described in {\it Materials and Methods}. All values are average values from duplicate repetitions. For a detailed discussion see main text.}
\end{figure}

To probe the transient response of the Cad module, we first grew {\it E. coli} (strain MG1665) to exponential phase at pH 7.6 in minimal medium. We then induced the Cad module by transferring cells into fresh minimal medium with 10 mM lysine and buffered at pH 5.8. The induction defined the starting point, $t=0$ min, for our measurements of the response, which we performed initially at intervals of 5 minutes, until $t=30$~min, and then at longer intervals of 30 minutes. To quantify the response, we assayed the {\it cadBA} mRNA level, the specific CadA activity, and the external cadaverine concentration, see {\it Materials and Methods} for all experimental details. The results are shown in Fig.~\ref{FIGtimeseries_exp}. Over the entire 4 hour period of the experiment, the external pH, shown in Fig.~\ref{FIGtimeseries_exp}~(a), remained low, even slightly decreasing from the induction level. Transcription of the {\it cad} operon began immediately after induction, and mRNA rapidly accumulated as shown in Fig.~\ref{FIGtimeseries_exp}~(b). At $t\approx 25$~min the mRNA level peaked and then rapidly decreased, reaching its low pre-induction level at about $t=90$~min. The response on the protein level, as quantified by the specific activity of CadA shown in Fig.~\ref{FIGtimeseries_exp}~(c), was slower, exhibiting a slight delay after induction and reaching a plateau level at $t\approx 60$~min, which was sustained over the time of the experiment. The activity of the Cad module led to the production and secretion of cadaverine, which accumulated in the medium as shown in Fig.~\ref{FIGtimeseries_exp}~(d). 

The transient expression of the Cad module shown in Fig.~\ref{FIGtimeseries_exp} is in qualitative agreement with previous induction experiments which studied the system in a less quantitative manner \cite{Neely_JBacteriol_96}. The biochemical mechanism for the transient behavior remained unclear, however. It was suggested that the external cadaverine level exerts a negative feedback on the activity of the regulator CadC \cite{Neely_JBacteriol_94,Neely_JBacteriol_96}. 
Alternatively, the Cad module might, for instance, directly affect and control the level of its input stimuli. In other words, the activity of the Cad module might reduce the external lysine concentration below its induction threshold, or shift the external pH level outside its range for induction.

\subsection*{Dose-response curves}

\begin{figure}
\centerline{\includegraphics[width=7cm]{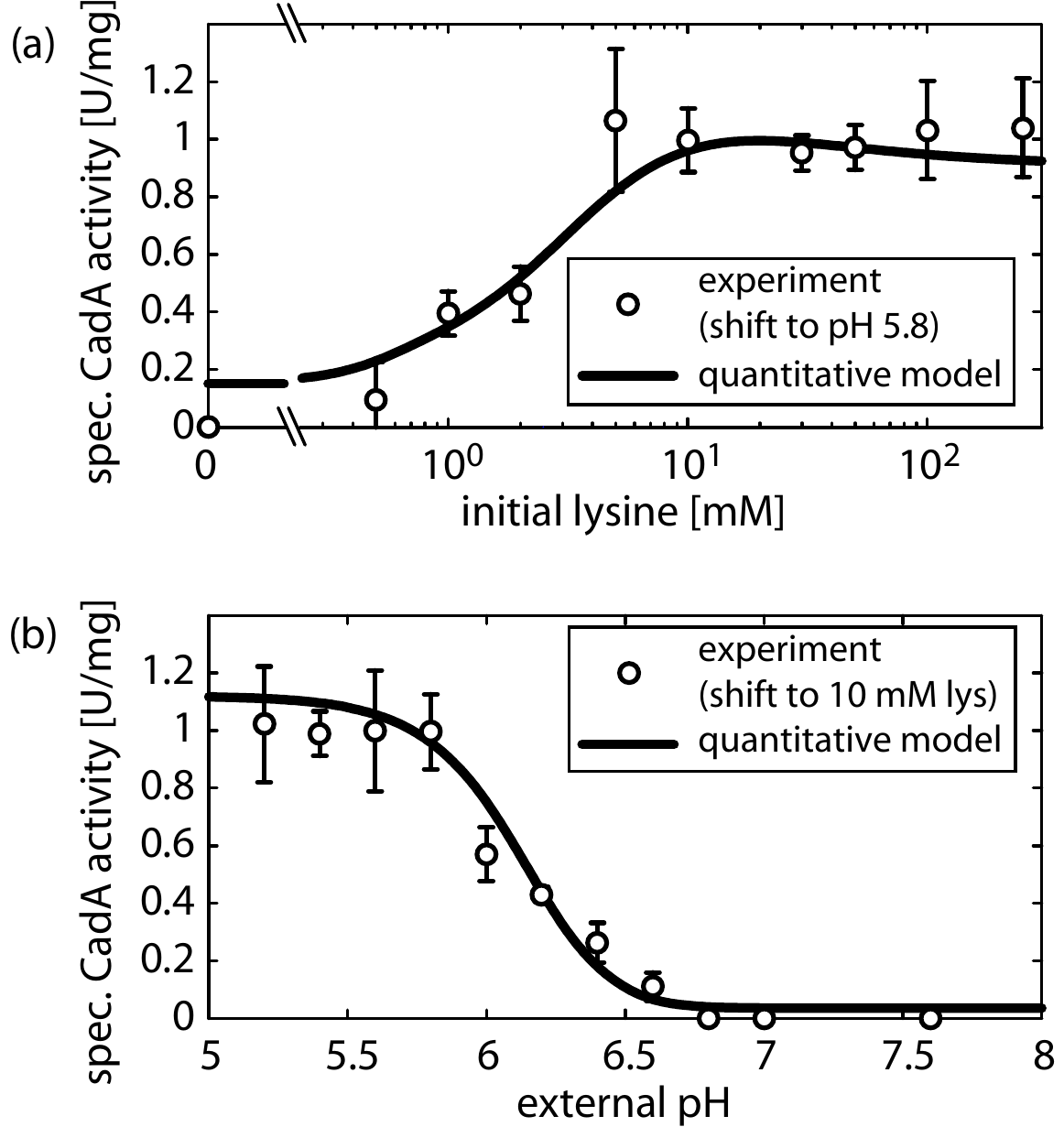}}
\caption{\label{FIGolson_data} Dose-response curves of the wild-type Cad system.
(a) Dependence of the long-term CadA response on the inducing lysine concentration. The specific CadA activity was determined 8 h after induction with pH 5.8 and stated lysine concentration, see {\it Materials and Methods}.  Similarly, the dependence on the inducing pH in (b) was obtained by induction with 10 mM lysine and stated pH. Here activity was determined 90 min after induction.
The solid lines show the fit result of our quantitative model.}
\end{figure}

To address these possible alternative explanations, and to characterize the ranges and the intensity of the total response under our experimental conditions, we next determined the ``dose-response'' behavior of the Cad module. We have seen above, see Fig.~\ref{FIGtimeseries_exp}~(c), that the CadA activity reaches a steady-state plateau about 60 min after induction. We take this plateau value as a proxy for the total (cumulative) response of the Cad system and study its dependence on the input signals. To this end, we induced the Cad module with different external pH levels and initial lysine concentrations, and assayed samples at least 90 min after induction for their CadA activity, see {\it Materials and Methods} for details. 
Fig.~\ref{FIGolson_data}~(a) and (b) show the lysine- and pH-dependence of the response, respectively. The data in Fig.~\ref{FIGolson_data}~(a) indicates that when induced with pH 5.8, the Cad module is barely active at lysine concentrations below 0.5 mM, whereas it is fully active for lysine levels exceeding 5 mM. In between these values, the activity increases sigmoidally with the inducing lysine concentration. Similarly, at a given lysine induction level of 10 mM, the activity depends sigmoidally on the inducing pH, see Fig.~\ref{FIGolson_data}~(b), with no significant activity above pH 6.8, and full activity at pH 5.8 and below. 

Taken together, the pH-dependence of the total response shown in Fig.~\ref{FIGolson_data}~(b) and the time series in Fig.~\ref{FIGtimeseries_exp}~(a) shows that the pH level did not leave the range for induction during the course of our experiment in Fig.~\ref{FIGtimeseries_exp}. Hence, the transient behavior of the {\it cadBA} expression is clearly not mediated by a decrease of the external pH stimulus. 

The dose-response curves of Fig.~\ref{FIGolson_data} characterize the input-output behavior of the system when the Cad module is regarded as a "black box" signal processing unit. In particular, we can read off the apparent activation thresholds of the Cad module, i.e., the pH and lysine levels at which the module displays half-maximal activity. Such apparent thresholds constitute the first level of description in a top-down system analysis. Conversely, in a bottom-up analysis, the first level of description is via biochemical interaction parameters, while effective parameters such as apparent thresholds emerge from the interplay of molecular interactions. Quantitative modeling and analysis of this interplay is the only way to connect between the two levels of description. In the following, we want to make such a connection, and then leverage it to estimate the molecular activation thresholds of CadC from the apparent behavior of the Cad module.

\subsection*{Construction of a quantitative model}

It is clear from the above that a minimal quantitative model of the Cad module must describe the integration of the input signals pH and lysine, as well as the effect of cadaverine on the activity of CadC. Furthermore it must describe the regulation and expression of the {\it cadBA} operon, and the functioning of the CadA and CadB proteins, such that we may relate the quantitative model to the observed dynamics of the three system variables monitored in our experiments. Our construction of such a quantitative model is guided by the qualitative model in Fig.~\ref{FIGscheme} and the known biochemistry of the Cad module. 

\paragraph*{Signal integration.} 

The three external signals known to affect the activity of the Cad module are the time-dependent lysine concentration, $l(t)$, cadaverine concentration, $c(t)$, and the pH. The membrane protein CadC, which receives and combines these signals into a single response \cite{Watson_JBacteriol_92, Neely_JBacteriol_94, Dell_MolMicrobiol_94}, is constitutively expressed \cite{Dell_MolMicrobiol_94}, and hence we take the total amount of CadC per cell, $C_{0}$, to be constant. The signals then modulate only the fraction of active CadC molecules per cell, $C(t) / C_{0}$. We assume that the different signals regulate CadC independently, such that the CadC activity is described by the product form
\begin{equation}
\label{EQinput0}
 C(t) / C_{0} =  f(\mathrm{pH}(t))\times g(l(t))\times h(c(t))\,.
\end{equation}
Indeed, all experimental data available show no indication for a coupled effect of the input signals on CadC \cite{Neely_JBacteriol_96, Tetsch_MolMicrobiol_08}. Another assumption implicitly made by Eq.~(\ref{EQinput0}) is that the fraction of active receptors is always equilibrated to the current levels of the input signals, i.e., it does not depend on the signal levels prior to the time $t$. This assumption is also plausible, since the typical timescale for conformational transitions in receptors (see, e.g., Ref.~\cite{Sourjik_PNAS_02}) is much shorter than the timescale of our experiments. The functions $f$, $g$ and $h$ in Eq.~(\ref{EQinput0}) take on values between 0 and 1, and are assumed to be of the Hill form typical for cooperative binding reactions. The pH-dependence is parameterized as 
\begin{equation}
  f(\mathrm{pH}) = \frac{1}{1+10^{\frac{\mathrm{pH}-\mathrm{pH_0}}{\Delta \mathrm{pH}}}}
  \label{pH-dep}
\end{equation}
with $\mathrm{pH_0}$ denoting the pH value at which $f$ reaches half-maximal activity and $\Delta \mathrm{pH}$ determining the width of the response curve. 
Similarly, the lysine and cadaverine dependence take the form 
\begin{equation}
  g(l) = \frac{(l/K_l)^{n_l}}{1+(l/K_l)^{n_l}}\,, \quad
  h(c) = \frac{1}{1+(c/K_c)^{n_c}} \;,
  \label{lys-cad-dep}
\end{equation}
where $K_l$ and $K_c$ are the effective {\it in vivo} activation thresholds for the direct and indirect regulatory interactions of lysine and cadaverine with CadC. As usual, the Hill coefficients, $n_l$  and $n_c$, parameterize the cooperativity of the binding reactions and determine the maximal sensitivity for signal detection. The difference in form between $g(l)$ and $h(c)$ stems from the fact that lysine activates the Cad module whereas cadaverine represses. On the other hand, the difference to Eq.~(\ref{pH-dep}) is because the pH is logarithmically related to the \Hp concentration.

\paragraph*{Transcriptional regulation.}  

In its activated conformation, CadC directly binds to the {\it cadBA} promoter and activates {\it cadBA} expression \cite{Kueper_JMolMicrobiolBiotechnol_05}. Generally, transcriptional regulation in bacteria can be described by quantitative ``thermodynamic'' models \cite{Bintu_CurrOpinGenetDev_05a, Bintu_CurrOpinGenetDev_05b}. For the present case, an appropriate form for the transcriptional activity from the $P_{cad}$ promoter as a function of the abundance of active CadC, $C(t)$, is derived in {\it Materials and Methods}. The resulting rate equation for the time evolution of the mRNA level $m$ then takes the form 
\begin{equation}
  \frac{d}{dt} m(t) = \nu_m \left( \frac{ 1+ (C(t)/K_{C})^2 f}{1+(C(t)/K_{C})^2} \right)^2 - \lambda_m\, m(t) \,,
  \label{EQmRNA}
\end{equation}
with the basal transcription rate $\nu_m$, the degradation rate $\lambda_m$, the fold-change $f$ between basal and maximal transcription rate, and $K_{C}$ denoting the binding threshold for CadC-DNA binding. The particular choice of the exponents in the first term is motivated by the observation that the {\it cadBA} promoter appears to be regulated by two binding sites for dimeric forms of CadC \cite{Kueper_JMolMicrobiolBiotechnol_05}.

\paragraph*{Kinetics of enzyme expression and catalysis.}  

On the protein level, we have a similar interplay of synthesis and decay as in Eq.~(\ref{EQmRNA}), 
\begin{equation}
  \frac{d}{dt} A(t) =  \nu_p\, m(t) - \lambda_p\, A(t)\,, \label{EQCadA} 
\end{equation}
where $A$ is the abundance of CadA per cell, and $\nu_p$ and $\lambda_p$ are the translation and degradation rate, respectively. The level of the transporter CadB is taken to be proportional to that of CadA. Since they are translated from the same mRNA, this means that we neglect possible post-transcriptional regulation, for which there seems to be no experimental indication \cite{Meng_JBacteriol_92b}. We also assume that we can subsume the transport and turnover of lysine to cadaverine through CadB and CadA by a single effective reaction, since little is known about the microscopic rates and affinities of the coupled transport and decarboxylase reactions. As detailed in {\it Materials and Methods}, this assumption leads us to
\begin{equation}
  \label{lys-turnover}
  \frac{d}{dt} l(t)  = - v_{max}A(t)\, \frac{\, l(t)}{K_{m} + l(t)}\,. 
\end{equation}
This simplified reaction corresponds to an effective Michaelis-Menten process with external lysine as the substrate, an effective maximal lysine turnover rate $v_{max}$, and an effective Michaelis constant $K_m$. Implicit in Eq.~(\ref{lys-turnover}) is also the assumption that growth of the bacterial population over the time period of the experiment is negligible. 
For the external cadaverine level, we assume the flux balance 
\begin{equation}
  \frac{d}{dt} c(t)  = - \frac{d}{dt} l(t)   \;,
  \label{EQcadaverine}
\end{equation}
implying that the sum of external lysine and cadaverine is conserved at all times, $l(t) + c(t) =\mathit{const}$. This flux balance appears justified, given experimental results with a LysP-deficient mutant strain which we report and discuss further below. 
We take the pH to be a constant over the duration of our kinetic experiments, since the pH changes only very little in our buffered medium, and the Cad module is not very sensitive to the pH over this regime (see above).

\begin{table}[b]
\caption{\label{TABcadaverine} 
Repression of the long-term Cad response by cadaverine (data taken from Neely {\it et al.} \cite{Neely_JBacteriol_94}). The {\it cadBA} expression ({\it central column}), as determined from the $\beta$-galactosidase activity of a $cadA-lacZ$ fusion, was measured in cells that were grown in medium pH 5.8 with 10 mM lysine and the indicated cadaverine concentrations for 3 h \cite{Neely_JBacteriol_94}. The model values in the right column show the fit result of our quantitative model.
}
\begin{center}
{\small {\sf 
\begin{tabular}{|r@{$\;\;$}|@{$\;\;$}c@{$\;\;$}|@{$\;\;$}c@{$\;\;$}|}
\hline
initial cadaverine  & rel. {\it cad} expression & model value \\
\hline 
0 $\mu$M & 1.00 & 0.97 \\
20 $\mu$M & 0.89 & 0.92 \\
80 $\mu$M & 0.60 & 0.74\\
320 $\mu$M & 0.12  & 0.12\\
1300 $\mu$M & 0 &  0.05\\
\hline
\end{tabular}
}}
\end{center}
\vspace*{0.2cm}
\end{table}

\subsection*{Data interpretation with the quantitative model}  

We will now demonstrate that the simple quantitative model constructed above is indeed a powerful tool. First, we test to what extent this model is compatible with the data sets reported above. For this test, we also include another data set from Neely {\it et al.} \cite{Neely_JBacteriol_94}, who determined the cadaverine-dependent dose-response of the Cad module, see Table~\ref{TABcadaverine}. This additional data further constrains our model and probes the consistency with the existing literature.

\begin{figure}[t]
\centerline{\includegraphics[width=8cm]{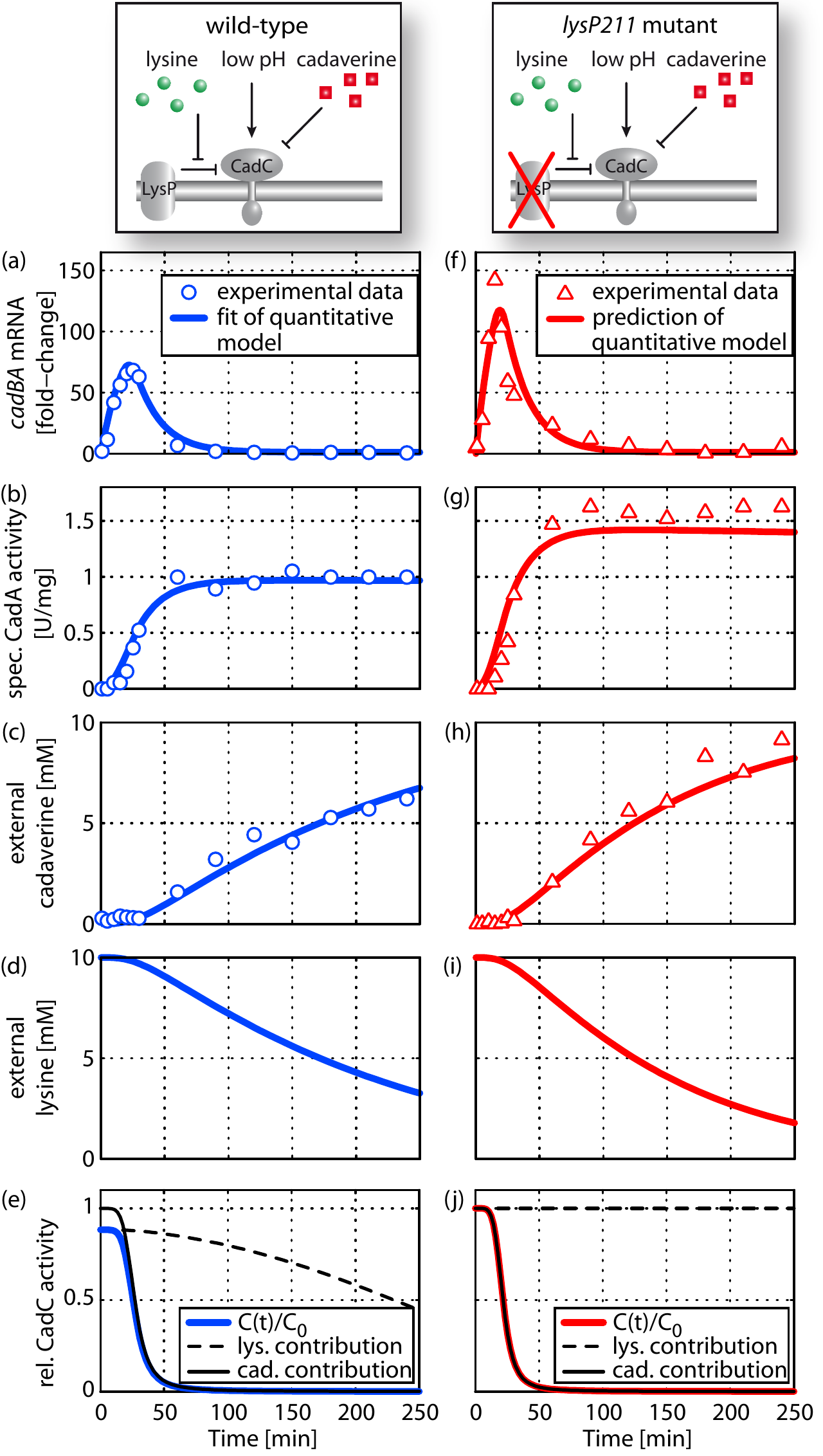}}
\caption{\label{FIGtimeseries} Fit of our quantitative model ({\it blue lines, left column}) to the experimental data of the wild-type induction kinetics ({\it blue circles}). The parameter-free prediction of our model for a LysP-deficient mutant  ({\it red lines, right column}) displays reasonable agreement with the experimental data ({\it red triangles}). The dashed and solid black lines in (e) and (j) show the lysine and cadaverine contribution ($g(l(t))$ and $h(c(t))$ in Eq.~(\ref{lys-cad-dep})) to the signal integration function in Eq.~(\ref{EQinput0}).}
\end{figure}

\begin{table*}
\caption{\label{TABparameters} Parameters of the quantitative model. \\ LB - lower bound; UB - upper bound; Estimated parameter values are shown as  $(\mathsf{best\, fit\, value})\pm {\mathsf{\sigma^+} \atop \mathsf{\sigma^-}}$, where $\sigma^+$ and $\sigma^-$ indicate the asymmetric standard errors in positive and negative direction, respectively, see Eq.~(\ref{EQNerrors}) in {\it Materials and Methods}.
}
\begin{center}
{\tiny {\sf
\begin{tabular}{l@{$\;$}p{5.5cm}@{$\;\;$}r@{$\;$}r@{$\;\;\;\;$}r@{$\,$}l|@{$\;\;$}p{3.5cm}}
\hline
& Parameter & LB & UB & Estimated& Value   & Comment \\
\hline \\
& {\bf  \textsf{Sensory module}} & & & & \\ 
$\mathsf{K_l}  $ & Threshold for CadC activation by lysine  & 1 & 20 &  $\mathsf{3.6}\pm {\mathsf{5.8} \atop \mathsf{0.6}}$\, & mM & bounds suggested by Fig.~\ref{FIGolson_data}~(a)\\
$\mathsf{K_c} $ & Threshold for CadC inactivation by cadaverine & 50 & 1000 & $\mathsf{235}\pm {\mathsf{32} \atop \mathsf{49}}$ & $\mu$M & bounds suggested by Fig.~\ref{FIGolson_data}~(a) \\
$\mathsf{n_l}  $ & Hill coefficient for CadC regulation by lysine & 1 & 5 &  $\mathsf{1.1}\pm {\mathsf{0.2} \atop \mathsf{0.1}}$ & & \\
$\mathsf{n_c}  $ & Hill coefficient for CadC regulation by cadaverine & 1 & 5 & $\mathsf{2.8}\pm {\mathsf{0.9} \atop \mathsf{0.3}}$ & & \\
$\mathsf{pH_0}  $ & pH threshold for CadC activation & - & - & 6.2 & & estimated from data in Fig.~\ref{FIGolson_data}~(b)\\
$\Delta \mathsf{pH}  $ & width of the transition from active to inactive CadC & - & - & 0.5 & & estimated from data in Fig.~\ref{FIGolson_data}~(b) \\
& &\\
& {\bf \textsf{Expression module}} & \\ 
$\mathsf{C_{0}/ K_{C}}$ & Total CadC per cell in relation to the threshold for CadC-promoter binding & 0.1 & 10 & $\mathsf{1.1}\pm {\mathsf{2.6} \atop \mathsf{0.1}}$  & &  The level of CadC is just sufficient to activate the pathway \cite{Watson_JBacteriol_92}, suggesting that {\it \textsf{in vivo}} $\mathsf{C_0/K_{C}\approx1}$.\\
$\mathsf{\nu_m}$ & basal transcription rate  & 0.001 & 0.1 & $\mathsf{4.3} \pm {\mathsf{14.1} \atop \mathsf{2.2}}$ & $\times \mathsf{10^{-3}}\, \mathsf{min}^{-1}$ & \\
$\mathsf{f}$ & fold-change between basal and maximal transcription rate & 10 & 1000 & $\mathsf{698}\pm {\mathsf{170} \atop \mathsf{452}}$ & & typical range \cite{Lutz_Nucl_Acids_Res_97,Kuhlman_PNAS_07}  \\ 
$\mathsf{\nu_p}$ & effective translation rate & $10^{-4}$ & $10^{-1}$ & $\mathsf{4.2} \pm {\mathsf{8.6} \atop \mathsf{2.3}}  $& $\times \mathsf{10^{-3}}$ U/min &  effective parameter with broad range \\
$\mathsf{\tau_m}$&   mRNA half-life $(= \mathsf{\ln 2/\lambda_m})$& 1 & 50 & $\mathsf{13.8}\pm {\mathsf{0.4} \atop \mathsf{1.2}}$ & min & typical range \cite{Bernstein_PNAS_02}\\
$\mathsf{\tau_p}$&  protein half-life $(= \mathsf{\ln 2/\lambda_p})$& 1 & 10$^\mathsf{4}$& $\mathsf{29}\pm {\mathsf{2137} \atop \mathsf{4}}$ & h & CadA is expected to be stable \cite{Sabo_Biochemistry_74}\\
$\mathsf{v_{max}}$& maximal rate for lysine turnover via CadA and CadB & 10$^\mathsf{-4}$ & $10$ & $\mathsf{1.3\pm {\mathsf{1.4} \atop \mathsf{0.5}}}$ & $\mathsf{\times 10^{-3}} \, \mathsf{min}^{-1}$ & effective parameter with broad range\\
$\mathsf{K_m}$ & effective Michaelis constant for lysine turnover via CadA and CadB & 1 & 100 & $\mathsf{26}\pm {\mathsf{37} \atop \mathsf{12}}$  & mM & effective parameter with broad range \\ \hline\\
\end{tabular}
}
}
\end{center}
\end{table*}

\begin{figure}
\centerline{\includegraphics[width=7cm]{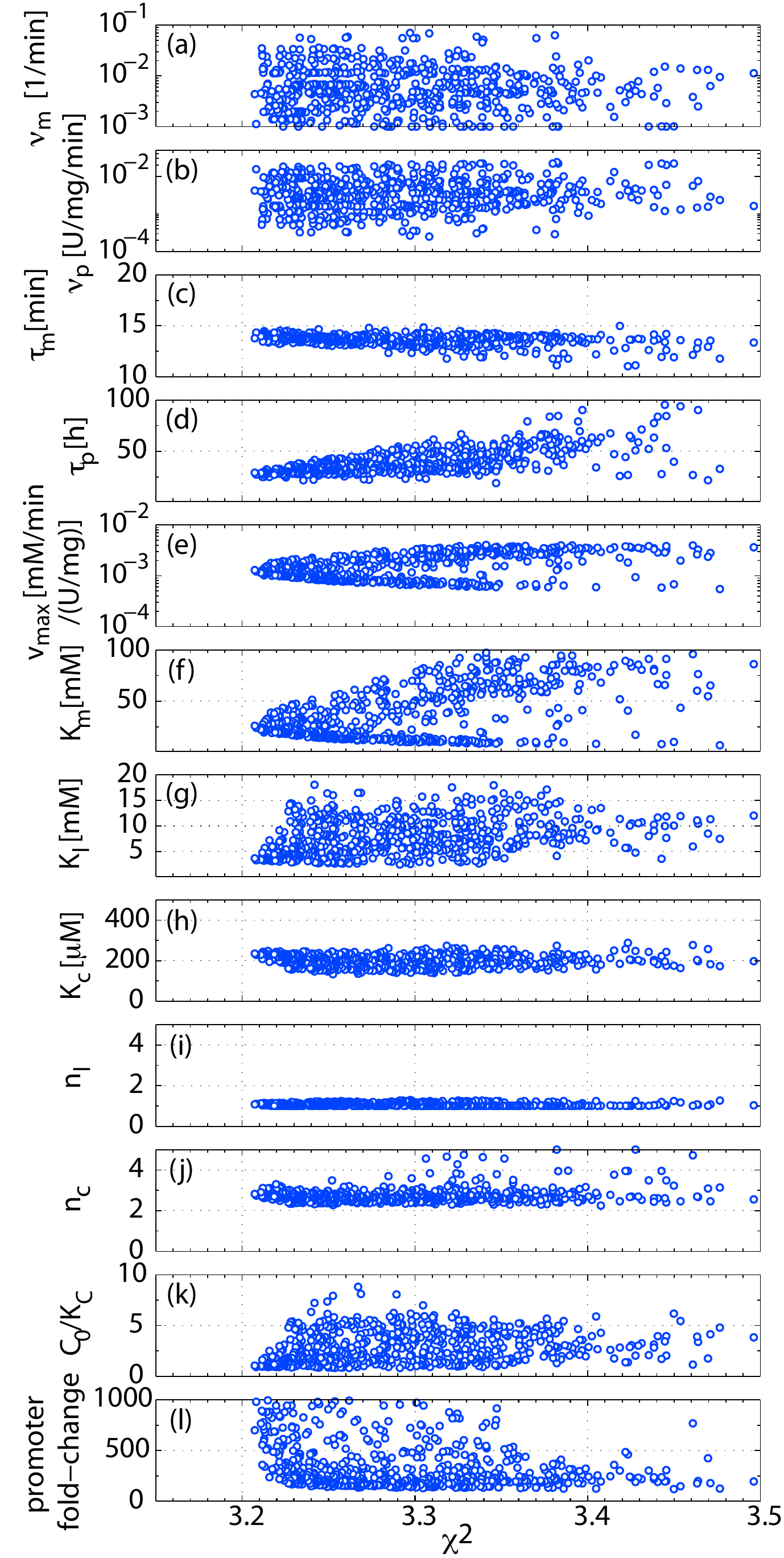}}
\caption{\label{FIGresidual_parameter_corr} Correlations between the goodness of fit and the estimated parameters. \\ The points correspond to local optima in the parameter space, for which the difference between the quantitative model and the experimental data in Figs.~\ref{FIGtimeseries_exp}, \ref{FIGolson_data} and Table~\ref{TABcadaverine} is minimized, see {\it Materials and Methods} for all details. As the fit quality increases (lower $\chi^2$) most parameters are confined to narrow intervals, indicating that their values are well constrained by the experimental data. However, some parameters display significant variation even for the lowest $\chi^2$ values and from parameter-parameter correlation analysis in Fig.~S1 one finds, that only combinations of those are well confined by our data.}
\end{figure}

In total, the quantitative model has 14 parameters. We constrained each of these to a range inferred from typical physiological values and other information in the literature, see Table~\ref{TABparameters}. We then fitted our model to all data sets simultaneously using standard least-squares minimization of the residual $\chi^2$, as described in {\it Materials and Methods}. The curves corresponding to the best fit parameters are shown in Figs.~\ref{FIGolson_data} and \ref{FIGtimeseries} (blue lines). The overall agreement with the experimental data is good, both for the response dynamics in Fig.~\ref{FIGtimeseries}~(a)-(c) as well as for the dose-response in Fig.~\ref{FIGolson_data} and Table~\ref{TABcadaverine}. 
Note however, that not all model parameters are individually well constrained by the data. This becomes apparent by plotting the correlations between the quality of the fit, characterized by the residual $\chi^2$, and the fit parameters, as shown in Fig.~\ref{FIGresidual_parameter_corr}. As the fit becomes better (lower $\chi^2$), most parameter values are confined to a narrow interval, indicating that the information contained in the experimental data accurately determines their values. For instance, the {\it cadBA} mRNA lifetime $\tau_m$ determines the decay time of the transient expression peak in Fig.~\ref{FIGtimeseries}~(a), and is therefore strongly constrained in our model. In contrast, some of the parameters display a wide variation even at the lowest $\chi^2$ values, e.g. the transcription and translation rates, $\nu_m$ and $\nu_p$. For these cases, where individual parameters are 'sloppy' \cite{Gutenkunst_PlosCompBiol_07}, certain combinations of these parameter are well constrained by the data sets. Pairwise scatterplots, as shown in Fig.~S1 of the {\it Supplementary Material}, identify correlations and anticorrelations between the parameters and help to reveal the appropriate combinations. For instance, the product of the transcription and translation rates is much better determined by the data than the individual rates. 

It is noteworthy that the best-fit value for the effective Hill coefficient $n_{c}$ for the regulation of CadC by cadaverine is close to 3, and is relatively well constrained by the data. This suggests that a molecular mechanism for cooperativity is at work, possibly a multimerization of CadC proteins in the membrane. Another interesting observation from Table~\ref{TABparameters} concerns the half-life of {\it cadBA} mRNA, which was well-constrained by the data to a value of almost 14 min. A global analysis of RNA half-lifes in {\it Escherichia coli} \cite{Selinger_GenomeRes_03} found an extremely short half-life of less than 2 min for the {\it cadBA} mRNA, suggesting that an active degradation mechanism is involved. Our Northern blot data for the {\it lysP211} mutant, shown in Fig.~\ref{FIGtimeseries}~(f), does indeed suggest a rapid decay of the mRNA at high levels directly after the peak, followed by a slower decay at lower levels. Our quantitative model only allows for a single degradation rate, which leads to the intermediate half-life of 14 min as a best-fite value. However, the changing degradation rate could be rationalized under the assumption that the {\it cadBA} mRNA has a relatively weak binding affinity to the degrading enzyme, such that active degradation only contributes significantly at high mRNA levels.

\begin{figure}
\centerline{\includegraphics[width=7cm]{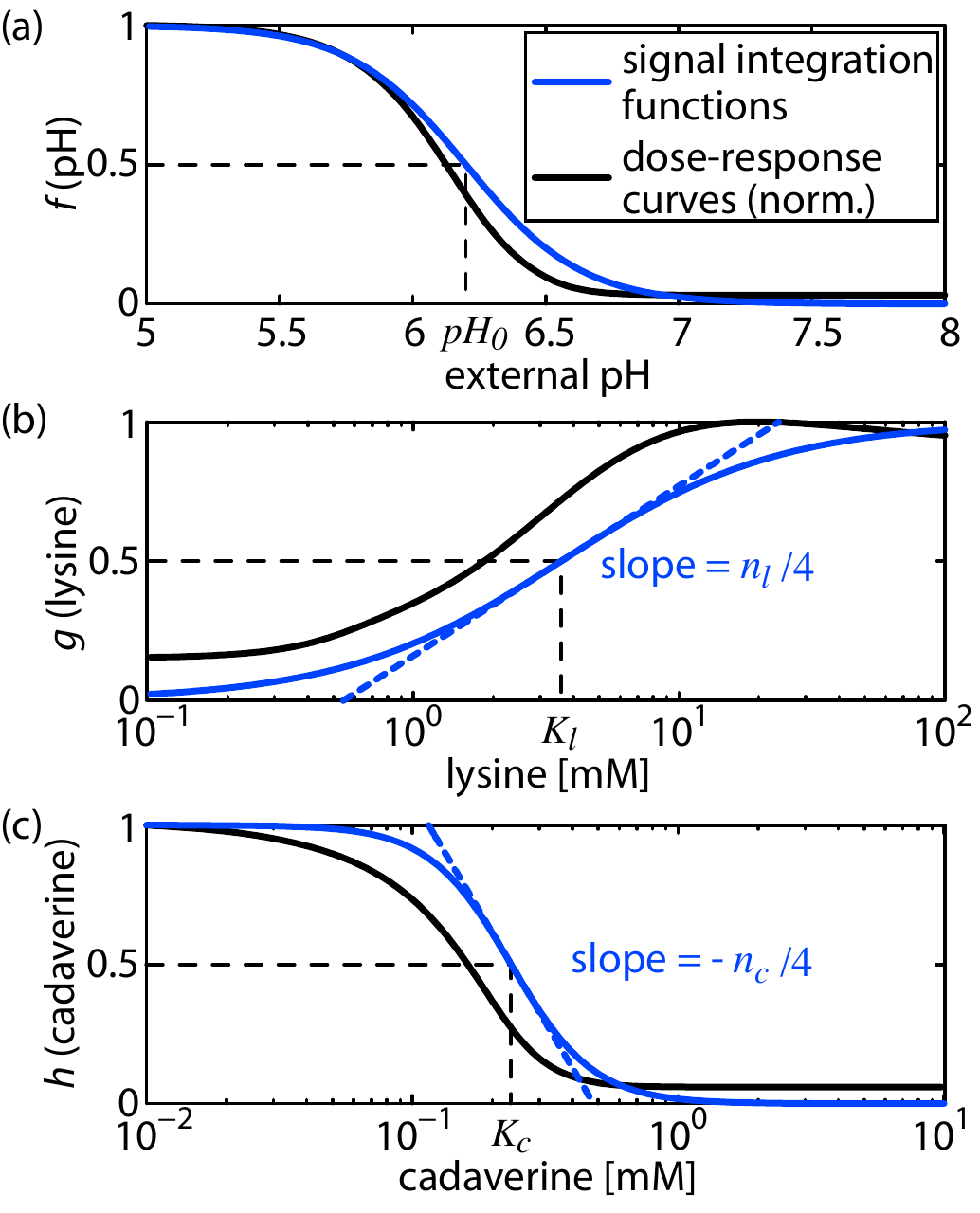}}
\caption{\label{FIGCadCActivity} Comparison of the extracted signal integration functions ({\it blue curves}) with the normalized dose-response curves ({\it black curves}). \\ The signal integration functions describe the dependency of the CadC activity on the pH level (a), on lysine (b) and on cadaverine (c). They correspond to the functions $f$, $g$ and $h$ in Eqs.~(\ref{pH-dep}) and (\ref{lys-cad-dep}) and are plotted for the best fit parameters listed in Table~\ref{TABparameters}. The apparent activation thresholds of the dose-response curves ({\it black curves}) are shifted with respect to the inferred biochemical thresholds of CadC ({\it blue curves}), since the dose-response curves also depend on the biochemical properties of the nonlinear feedback in the Cad module. From the extracted signal integration functions we can also read off the sensitivity of CadC on its input signals, since their maximal slopes are determined by the Hill coefficients $n_l$ and $n_c$.}
\end{figure}

Given the compatibility of model and data, we next used the model to infer quantitative characteristics of the module that are not directly assayed. For instance, the activity of the central regulator CadC as a function of the external signals lysine, pH, and cadaverine is a biochemical characteristic that is pivotal to the function of the module, but difficult to measure directly. In our model, this quantitative characteristic is represented by the signal integration function, Eq.~(\ref{EQinput0}). On the other hand, the experimental data shown in Fig.~\ref{FIGolson_data} probes the final output of the Cad module on the protein level (specific CadA activity), which is the integrated result of a nonlinear dynamical system with feedback. 
Fig.~\ref{FIGCadCActivity} compares this final system output (black curves, as in Fig.~\ref{FIGolson_data}, but normalized to one) with the inferred activity of the CadC regulator (blue curves). The latter represent the three sigmoidal functions that make up the signal integration function (\ref{EQinput0}), with the parameters determined from the global fitting procedure described above. 
We observe from Fig.~\ref{FIGCadCActivity} that the final system output behaves qualitatively similar to the inferred biochemical activity of CadC. However, in each case, the apparent activation threshold for the system response (point of half-maximal CadA activity, black curve) is shifted with respect to the inferred biochemical activation threshold (point of half-maximal CadC activity, blue curve). These shifts are due to the fact that the total CadA activity does not only depend on the characteristics of the regulator, but also on the biochemical properties and timescales of the negative feedback loop. In principle, the feedback can even lead to non-monotonic behavior in the dose-response, despite the underlying monotonic dependence of the signal integration function on the levels of the external signals\footnote{The model predicts indeed a weak non-monotonic effect in the lysine dependence, however this is not a robust prediction.}. The inferred {\it in vivo} biochemical activation thresholds, $K_l = 3.6$\,mM and $K_c = 235\,\mu$M, and the values of the Hill coefficients describing the sensitivities to the signals, can be read off directly from the blue curves, as indicated in Fig.~\ref{FIGCadCActivity}. 

It is useful to compare these biochemical activation thresholds to the actual concentrations encountered in our induction kinetics experiments, in particular at the time of transcriptional down-regulation ($t = 30$\,min). From the plots in Fig.~\ref{FIGtimeseries} (c) and (d) we see that the lysine concentration at this time point is still about a factor of 3 higher than the activation threshold, whereas the cadaverine concentration of $300\,\mu$M exceeds the deactivation threshold. The individual regulatory contributions from lysine and cadaverine to the CadC activity, i.e. $g(l(t))$ and $h(c(t))$ are plotted in Fig.~\ref{FIGtimeseries}~(e). The lysine curve (dashed line) displays only a very weak impact on the CadC activity, whereas the increase of cadaverine is the primary effect causing the down-regulation of the CadC activity (solid line). Hence, the analysis with our quantitative model strongly suggests that the negative feedback via external cadaverine can quantitatively explain the timing of the transient response in the wild-type Cad system and that the decreasing lysine stimulus is not involved in this behavior.

\subsection*{Prediction and experimental analysis under altered conditions}

So far we have analyzed the Cad module only in the wild-type strain and only with a single induction protocol. To obtain a more complete picture of its quantitative behavior, we constructed a mutant strain  MG1655-{\it lysP211} with a truncated and inactive form of the lysine permease LysP, see {\it Materials and Methods} for details. Qualitatively, we expected that this mutation would completely abolish the lysine requirement for the activation of CadC, since the truncated form of LysP would be unable to repress CadC \cite{Tetsch_MolMicrobiol_08} (an early study also indicated a de-repressed activation of the Cad system by a spontaneous mutation in {\it lysP} \cite{Popkin_JBacteriol_80}). 
Within our quantitative model, the {\it lysP211}  mutation was mimicked by setting the lysine-dependent activity function in Eq.(\ref{EQinput0}) equal to its maximal value, i.e., $g(l) = 1$, thereby rendering it independent of the inducing lysine concentration. Apart from this ``{\it in silico} mutation'', we left the model and the parameter values unchanged. 

We verified that the LysP-deficient mutant was indeed inducible by a shift from pH 7.6 to 5.8 alone, and did not require lysine for its induction (data not shown). Then, we performed the same induction kinetics experiments as with the wild-type strain, see the results in Fig.~\ref{FIGtimeseries} for the {\it lysP211} mutant (red triangles) and the quantitative model (red curves). We clearly see that the expression in the mutant remains transient, supporting again the conclusion that a signal different from lysine mediates the transcriptional down-regulation. The shape of the response is altered, however, with a stronger initial expression and a peak in the mRNA level that has a larger amplitude and reaches its maximum earlier than for the wild-type strain. Exactly these features are expected also on the basis of the quantitative model: The stronger initial expression is due to the full relief of the LysP repression, and the negative feedback via cadaverine sets in at an earlier time since the cadaverine threshold is more rapidly reached, see Figs.~\ref{FIGtimeseries}~(h) and (j). Also, the CadA activity, shown in Fig.~\ref{FIGtimeseries}~(g), is expected to reach a higher steady state plateau, as observed in the experiment. Interestingly, at the end of the experiment, the cadaverine level in Fig.~\ref{FIGtimeseries}~(h) reaches almost the 10~mM level of initially added lysine, which is in line with our flux balance assumption in the model (see previous section). 

\begin{figure}
\centerline{\includegraphics[width=7cm]{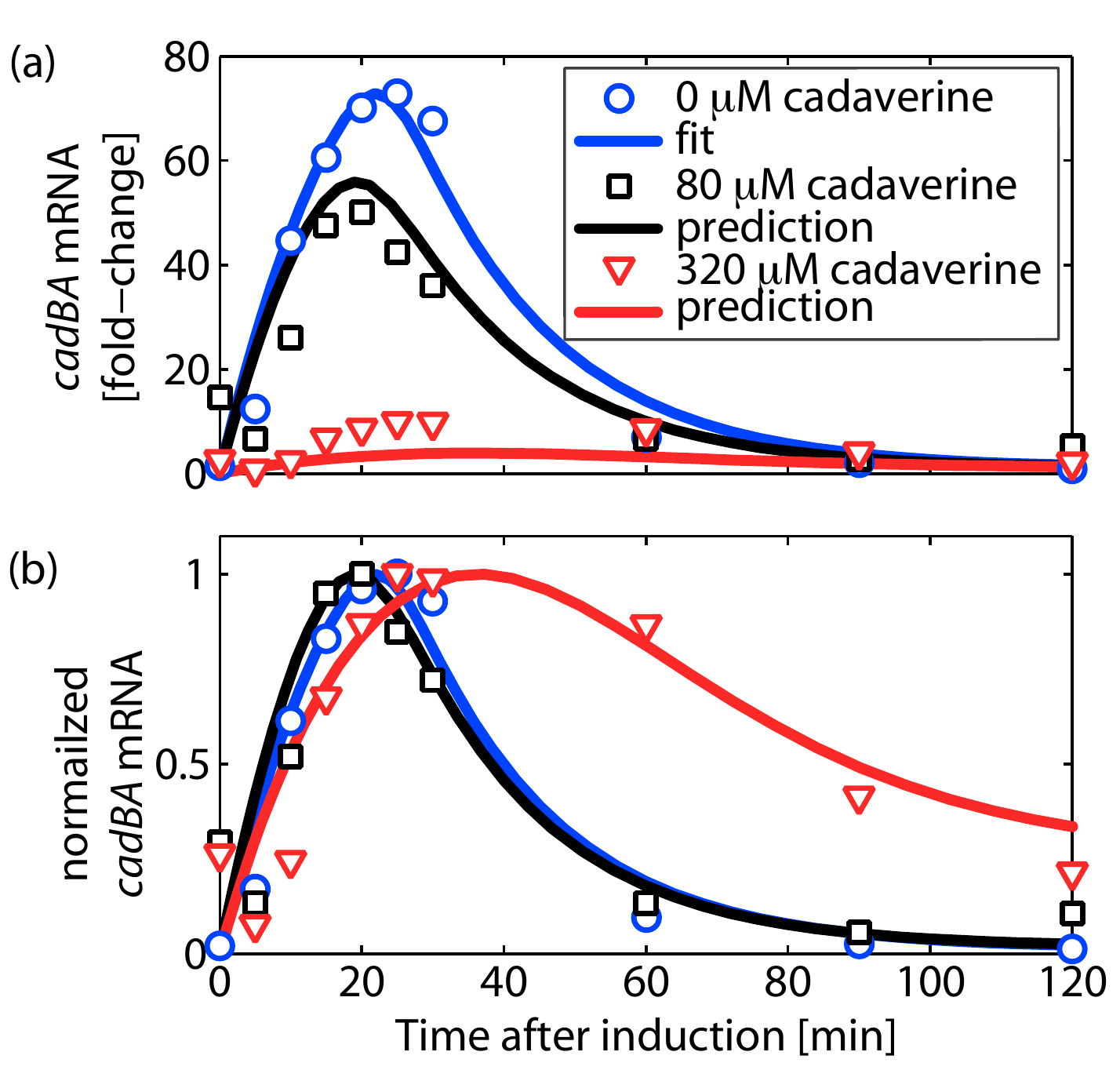}}
\caption{\label{FIGcadaverine_dependence} Experimental test of the kinetic model prediction. \\ The lines show the parameter-free model prediction for induction of the Cad system by a shift to pH 5.8 and $10$\,mM lysine, together with the cadaverine concentration indicated in the legend. The experimental data in (a) ({\it symbols}) was recorded and scaled as described in {\it Materials and Methods}. In (b) all data were normalized to their maximal values.}
\end{figure}

Next, we considered altered environmental conditions for the induction of the Cad module. Neely {\it et al.} \cite{Neely_JBacteriol_94} had already shown that induction with initially added cadaverine causes a significantly reduced long-term activity of the {\it cad} operon, cf. Table~\ref{TABcadaverine}. Yet, it is not clear whether this diminished long-term activity is caused by a {\it cadBA} expression pulse of similar strength but reduced duration, by a pulse of reduced strength with similar duration, or by a combination of both. To resolve this question, we again performed kinetic induction experiments with the wild-type strain, under identical conditions as in Fig.~\ref{FIGtimeseries_exp}, but additionally with $80\,\mu$M or $320\, \mu$M cadaverine supplied at the time of induction. In the latter case, the initially supplied cadaverine already slightly exceeded the inferred inactivation threshold of $K_c = 235\,\mu$M, such that a strong effect on the response could be expected. The resulting data is shown in Fig.~\ref{FIGcadaverine_dependence}~(a) (squares and triangles), together with the original data (no added cadaverine, circles) for comparison. Fig.~\ref{FIGcadaverine_dependence}~(b) shows the same data, but with all curves normalized to peak height 1, in order to emphasize the shape of the response. We observe that the primary effect of the addition of initial cadaverine is to reduce the strength of the response. This is also predicted by the quantitative model (solid lines), rather accurately for the $80\,\mu$M cadaverine data set, while the reduction for $320\, \mu$M cadaverine is predicted to be stronger than observed experimentally. 

It should be noted, that transcription was detected by Northern blot analysis, so that the shape of the response could be more accurately determined than the absolute amplitude. It is then interesting to observe that the quantitative model predicts a widening of the expression peak and a shift of the maximum to a later time for the highest cadaverine concentration ($320\, \mu$M, red curve). Again, this behavior results from the nonlinear feedback in the Cad module, i.e. the addition of initial cadaverine decreases the initial rate of mRNA production such that the CadA level increases more slowly, and the negative feedback also sets in more slowly. This predicted change in the shape of the expression peak for $320\, \mu$M agrees remarkably well with the experimental observation, see Fig.~\ref{FIGcadaverine_dependence}~(b). This finding provides strong evidence that the quantitative characteristics of the Cad module are well described and understood with the help of our mathematical model.

\section*{Discussion}

\subsection*{Conditional stress response with feedback inhibition}

In this work we analyzed the kinetics of a conditional pH stress response system, the lysine-decarboxylase system of {\it E. coli}, which exhibits only transient induction, even when the pH stress persists. Our results strongly suggest that the additional stimulus for the conditional response, i.e. a lysine-rich environment, is also not responsible for the transient behavior. Rather, our kinetic and dose-response experiments in combination with our quantitative model clearly indicate that a negative feedback via the product of the decarboxylation reaction, cadaverine, leads to the down-regulation of the response. 

Cadaverine has previously been linked to the transient behavior, in pivotal work on this stress response system \cite{Neely_JBacteriol_94, Neely_JBacteriol_96}. This link was based on the observations that external addition of cadaverine significantly reduces the long-term Cad activity and that a $\mathrm{CadA^-}$ mutant displayed persistent {\it cadBA} expression. However, if the transcriptional shut-off was mediated by a decrease of the lysine stimulus, the persistent {\it cadBA} expression could alternatively be explained by the lack of lysine consumption in the $\mathrm{CadA^-}$ mutant. And while a reduction in the steady-state activity implies that the Cad system is generally repressed by cadaverine, it is not clear that the timing of its down-regulation is set via this negative regulatory interaction. None of the previous studies directly measured the system-induced dynamics of the cadaverine concentration or studied the kinetics of the Cad system with externally added cadaverine. By performing these quantitative experiments with a high time resolution and by interpreting them with a quantitative theoretical model, we obtained evidence for a causal relation between the time of transcriptional down-regulation and the increase of the external cadaverine concentration above its deactivation threshold. For instance, the decreasing amplitude in the dynamical response after adding external cadaverine, as well as a more subtle delayed down-regulation, both predicted by the quantitative model, were strikingly confirmed by our kinetic measurements.

\subsection*{Signal transduction mechanism}

Our systems-level study of the Cad module also permits some conclusions about the involved molecular interactions and the signal transduction mechanism. Using our quantitative model, we were able to estimate the relevant {\it in vivo} activation and deactivation thresholds of the regulatory protein CadC. For the inactivation of CadC by external cadaverine we found a threshold of $K_c = 235\,\mu$M. This value is surprisingly close to the {\it in vitro} binding constant of $96\,\mu$M for the interaction of cadaverine with the periplasmic domain of CadC \cite{Tetsch_MolMicrobiol_08}. 
In contrast, CadC has almost no affinity for lysine. There is recent evidence that CadC is inhibited at low lysine concentrations via a transmembrane domain interaction with the lysine permease LysP, whereas the interaction is released at high lysine levels \cite{Tetsch_MolMicrobiol_08}. In the present work we determined the effective {\it in vivo} lysine activation threshold to be $K_l = 3.6$\,mM. This result is somewhat surprising, since the Michaelis constant $K_M$ for lysine transport by LysP is much lower at $\sim10 \, \mu$M  \cite{Rosen_JBiolChem_71}. However, the LysP/CadC interaction and the $K_M$ of LysP do not necessarily need to have a direct correspondence. 
	
An interesting open question concerns the signal transduction mechanism of CadC. Two alternative models have been proposed: 
(i) A reversible conformational transition of CadC activates its cytoplasmic N-terminal domain and allows it to bind to the promoter while remaining integrated in the membrane. 
(ii) A shift to acidic pH induces cleavage of the cytoplasmic domain, allowing it to diffuse freely to the promoter. 
The existence of the negative feedback by external cadaverine puts causal constraints on these microscopic mechanisms. If the cleavage mechanism were realized, it is not clear how external cadaverine could down-regulate {\it cadBA} expression after it has been induced, since the freely-diffusing cytoplasmic domains would no longer be able to recognize this signal. A high turnover of cleaved CadC could solve this problem, by rapidly eliminating unresponsive activators. However, the increased degradation of CadC under inducing conditions would have to be balanced by an elevated {\it cadC} expression. This, however, is in disagreement with previous observations, where it was found that {\it cadC}  expression is constitutive \cite{Watson_JBacteriol_92, Dell_MolMicrobiol_94}. Hence, the existence of the negative feedback by external cadaverine leads us to favor the reversible model.

\subsection*{Top-down system analysis of a functional module}

The Cad system of {\it E. coli} is a functional module with few closely connected molecular components. How such modules integrate, process, and respond to external signals is a central question, but generally also a difficult one. The approach taken in the current study is akin to a ``top-down'' system analysis, where input signals are controlled and the output(s), as well as key internal system variables, are measured. This is in contrast to a biochemical ``bottom-up'' approach, where each component would first be characterized separately and then their pairwise interactions, gradually moving upwards in complexity. In our study, we were able to bring these two complementary approaches into a first contact for the Cad module, with the help of our quantitative model which provided the means to estimate relevant {\it in vivo} values for biochemical interaction parameters and the quantitative form of the signal integration function displayed in Fig.~\ref{FIGCadCActivity}. Without the quantitative model, we would have been unable to extract this ``hidden information'' from the experimental data. Direct {\it in vivo} measurements of signal integration functions have been performed for the regulatory circuit controling chemotaxis in {\it E. coli}, using sophisticated single-molecule techniques. Since these powerful techniques are not easily transferred to the large class of functional modules of interest, the indirect approach taken in the present work may often be a welcome alternative.

\subsection*{Conclusion and Outlook}

Our quantitative analysis of the Cad system provides a first step towards a system-level understanding of the complex acid stress response network in microbes. Analogous quantitative studies of the two other major amino-acid decarboxylase systems, glutamate and arginine decarboxylase \cite{Foster_NatRevMicrobiol_04}, could reveal important insights in how these modules are orchestrated in the complex environment of its host. The presence of multiple amino acids in the natural environment of {\it E. coli} suggests that these conditional stress-response systems are often induced in parallel. It will be interesting to study how these systems are coordinated to provide an effective and robust way of pH homeostasis and acid tolerance response.

\section*{Materials and Methods}

\paragraph*{Bacterial strains and growth conditions.}
{\it E. coli} MG1655 \cite{Blattner_Science_97} was used as wild-type strain. The {\it lysP211}  mutant was obtained via undirected mutagenesis \cite{Popkin_JBacteriol_80,Tabor_JBacteriol_80}. For this purpose cells of {\it E. coli} MG1655 were grown on minimal agar plates with 0.2\% (w/v) glucose as sole carbon source \cite{Miller_ColdSpringHarbor_92} containing 100 $\mu$g/ml thiosine (S-aminoethyl cysteine), a toxic lysine analog which leads to spontaneous mutations in {\it lysP}. One mutant, designated MG1655-{\it lysP211}, had a nucleotide exchange at position 211 in {\it lysP} resulting in a stop codon and hence a truncated and inactive form of LysP (70 amino acids). 
{\it E. coli} strains MG1655 and MG1655-{\it lysP211} were grown aerobically in shaking flasks under non-inducing conditions at pH 7.6 in 5 L phosphate buffered minimal medium \cite{Epstein_JBacteriol_71} containing 0.4\% (w/v) glucose as sole carbon source at 37\celsius\,  to an $\mathrm{OD}_{600} = 0.5$ (non-inducing conditions). Subsequently, cells were collected by centrifugation (10 min, 4000\,g at 37\celsius), and transferred into fresh prewarmed minimal medium, pH 5.8 containing 10 mM L-lysine (L-lysine-hydrochloride, Roth) and 0.4\% (w/v) glucose in a 5 L fermenter (Biostat B, Satorius BBI Systems GmbH) (inducing conditions). Cultivation of cells was continued anaerobically at 37\celsius. At the indicated times, samples were taken, centrifuged at 4000\,g (4\celsius) for 5 min, and cell pellets as well as supernatants were separately stored at -80\textcelsius\, until further use. The number of colony forming units was determined after incubation of 100 $\mu$L of various dilutions on LB agar plates overnight at 37\celsius\, \cite{Sambrock_ColdSpringHarbor_89}.

\paragraph*{Lysine Decarboxylase Assay.}
Specific activity of the lysine decarboxylase CadA was measured by resuspending cells corresponding to 10 mL culture in 1 mL Ldc buffer (100 mM Na-acetate, pH 6.0, 1 mM EDTA, 0.1 mM pyridoxal phosphate, 10 mM $\beta$-mercaptoethanol, 10\% (w/v) glycerol). Lysozyme to 0.1 mg/mL was added, and the mixture was incubated on ice for 30 min. The cell suspension was sonified (10-20\% amplitude; 0.5 sec output puls; digital Branson Sonifier II 250), and the lysate was centrifuged at 15000\,g at 4\celsius\, for 15 min. Activity of lysine decarboxylase in cell-free extracts was measured as described \cite{Lemonnier_Microbiology_98} using 5\,$\mu$g protein per assay. Specific activity is defined as 1\,U/mg\,$ = 1\,\mu$mol cadaverine/(min $\times$ mg protein).

\paragraph*{Measurement of extracellular cadaverine.}

The extracellular cadaverine concentration was determined according to the spectrophotometric method described by Phan {\it et al.} \cite{Phan_AnalBiochem_82}. Briefly, 10 $\mu$L of culture supernatant was diluted 5 fold with $\mathrm{H_2O_{dest}}$, then 120 $\mu$L $\mathrm{Na_2CO_3}$ (1 M) and 120 $\mu$L TNBS (2,4,6-trinitrobenzene-sulfonic acid; 10 mM) were added, and the mixture was incubated for 4 min at 40\celsius. After extraction with 1 mL toluene, the absorption of the organic phase (containing N,N«-bistrinitrophenyl-cadaverine) at 340 nm was measured. The cadaverine concentration was calculated based on a standard curve using cadaverine-dihydrochloride (Sigma) between 0 and 500 nmol.

\paragraph*{Preparation of RNA.}

Total RNA was isolated according to the method of Aiba {\it et al.} \cite{Aiba_JBacteriol_81}. Briefly, cells were resuspended in cold 20 mM Tris/HCl, pH 8.0, and subsequently lysed by addition of 20 mM sodium acetate, pH 5.5, 0.5\% (w/v) SDS, 1 mM EDTA, pH 8.0. Then, RNA was extracted with prewarmed (60\celsius) acid phenol, and the mixture was centrifuged at 12000 g. After an additional extraction of RNA using phenol/chloroform/isoamylalcohol (25:24:1), RNA was precipitated with 100\% ethanol at -20\celsius\,  overnight. The precipitate was washed with 70\% (v/v) ethanol, and the dry RNA pellet was dissolved in 35 $\mu$L $\mathrm{H_2O_{dest}}$. RNA concentration was determined by measuring the absorption at 260 nm. All solutions were prepared with 0.1\% (v/v) DEPC (diethylpyrocarbonate).

\paragraph*{Northern blot analysis.}

Northern blot analysis was performed following the protocol described earlier \cite{Jung_JBacteriol_01}. Briefly, 20 $\mu$g RNA was separated by electrophoresis in 1.2\% (w/v) agarose-1.1\% (v/v) formaldehyde gels in MOPS (morpholinepropanesulfonic acid) buffer. RNA was transferred to a Hybond-Nylon membrane (GE Healthcare) by capillar blotting. Hybridization was performed following a standard protocol \cite{Sambrock_ColdSpringHarbor_89} using a [$\alpha^{32}$-P]dCTP labelled PCR fragment of the first 400 bp of cadBA. Radioactive labelling was quantified with a Phosphoimager. As control, expression of {\it rpoD}, a house-keeping gene of {\it E. coli} was analyzed. Signal intensity of {\it cadBA} mRNA was normalized to the signal intensity of {\it rpoD} mRNA. If not indicated otherwise, the data are given as fold-change of {\it cadBA} transcription relative to the pre-induction value. Additionally, in Figs.~\ref{FIGtimeseries} and \ref{FIGcadaverine_dependence} the absolute magnitude of the mRNA fold-change was rescaled, such that the integral over the expression curve was proportional to the long-term CadA activity.

\paragraph*{Model details.}

From thermodynamic models of transcriptional regulation reviewed in \cite{Bintu_CurrOpinGenetDev_05a,Bintu_CurrOpinGenetDev_05b}, the effective transcription rate, $\nu_m^\mathrm{eff}$, as a function of two activators $A$ and $B$ with independent binding sites (binding constants $K_A$ and $K_B$) is given by
\begin{equation}
\nu_m^\mathrm{eff} = \nu_m  \left( \frac{ 1+ (A/K_{A}) f_A}{1+(A/K_{A})} \right) \left( \frac{ 1+ (B/K_{B}) f_B}{1+(B/K_{B})} \right)\;,
\end{equation}
cf. ref.~\cite{Bintu_CurrOpinGenetDev_05a} Table~1, Case 10.
We make the simplifying assumption that both binding sites are identical and that each site can only be bound by a dimer of CadC ($C_2$) \cite{Kueper_JMolMicrobiolBiotechnol_05}. Setting $A=B=C_2$, exploiting mass action $K={C}^2/C_2$ and introducing the effective binding constant $K_{C}=\sqrt{K K_{C_2}}$ leads to the first term in Eq.~(\ref{EQmRNA}).

\begin{figure}[t]
\centerline{\includegraphics[width=8cm]{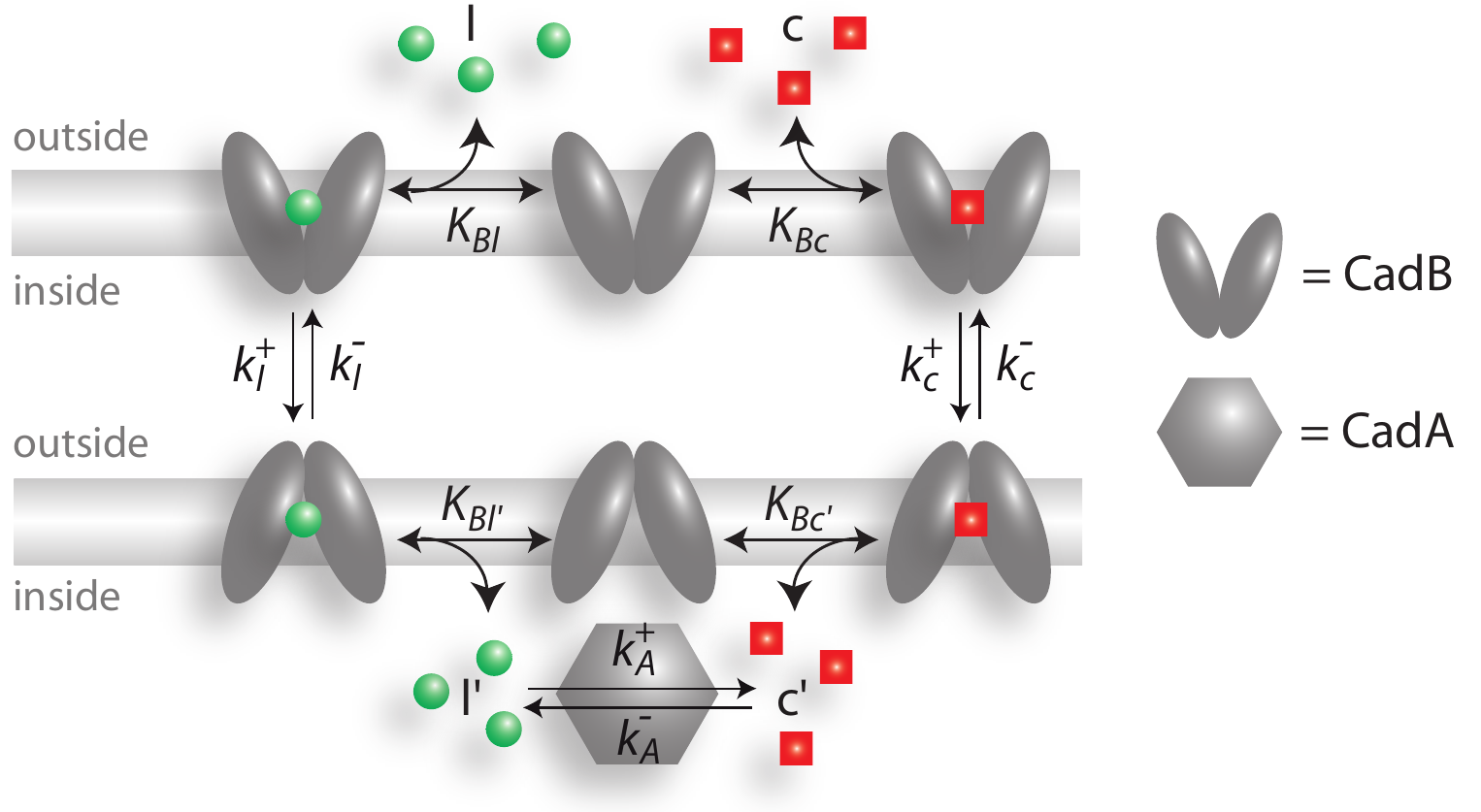}}
\caption{\label{FIGantiporter} Kinetic scheme of lysine and cadaverine transport and turnover. Antiport of lysine ($l$) and cadaverine ($c$) by CadB is modeled in analogy to the homologous arginine-ornithine antiporter ArcD in {\it L. lactis} by a single site Ping Pong Bi-Bi mechanism \cite{Driessen_JBiolChem_89}. Interconversion of lysine and cadaverine by CadA is modeled by a reversible first order reaction \cite{Koppelman_JBiolChem_58}. The prime (') indicates internal quantities. The rates ($k_l^+,k_l^-,k_c^+,k_c^-,k_A^+,k_A^-$) and equilibrium constants ($K_{Bl},K_{Bl'},K_{Bc},K_{Bc'}$) are indicated next to the reaction steps.
}
\end{figure}

The effective Michaelis-Menten form of the lysine turnover rate in Eq.~(\ref{lys-turnover}) was derived as follows. The transport of lysine and cadaverine via CadB was modeled in analogy to the homologous arginine-ornithine antiporter ArcD in {\it L. lactis} \cite{Driessen_JBiolChem_89}, see Fig.~\ref{FIGantiporter}. For low external cadaverine and internal lysine concentrations a general form the inwardly directed flux of lysine is \cite{Driessen_JBiolChem_89}
\begin{equation}
\label{EQNantiport_flux}
v_{eff} = k_l^+  B  \frac{[l]}{K_{Bl}+\left( 1+\frac{k_l^+}{k_c^-} + \frac{k_l^+ K_{Bc'}}{k_c^- [c']}\right) [l]}\,,
\end{equation}
where $B$ is the number of CadB molecules per cell and $[l]$ and $[c]$ are the concentrations of lysine and cadaverine on the outer surface of the membrane, respectively. Internal solute concentrations are marked by a prime ($'$). The parameters $k_l^+,k_c^-,K_{Bl}$ and $K_{Bc'}$ are defined in the kinetic scheme of Fig.~\ref{FIGantiporter}. Here it is assumed that (i) the conformational transition in CadB, which mediates the transport, does not occur without bound lysine or cadaverine, (ii) the antiporter and its substrates  are in binding equilibrium at each surface of the membrane, and (iii) the membrane translocation reaction of the carrier is slow and rate-limiting. The internal cadaverine concentration $[c']$ is determined by an interplay of the (reversible) decarboxylation  through CadA \cite{Koppelman_JBiolChem_58} and by the export via CadB. In steady state we find
\begin{equation}
[c'] = \frac{k_A^+}{k_A^-} [l'] - \frac{v_{eff}}{k_A^- A}\,,
\end{equation}
where $A$ is the number of CadA molecules per cell. If the equilibration between lysine and cadaverine through CadA is fast compared to the transport through CadB, the second term is negligible and the internal cadaverine level is solely determined by the internal lysine level $[l']$. In steady state it turns out that $[l']$ is not affected by the Cad module, since the 1:1 stoichiometry of the antiporter assures that lysine decarboxylation is balanced by lysine import.  If we further take advantage of the fact that CadB and CadA are transcribed polycystronically and that there seems to be no post-transcriptional regulation \cite{Meng_JBacteriol_92b}, we can set CadB proportional to CadA, i.e., $B = \alpha A$. Taken together,  in the limit of rapid CadA kinetics and low external cadaverine and internal lysine concentrations Eq.~(\ref{EQNantiport_flux}) reduces to the simple effective Michaelis-Menten form in Eq.~(\ref{lys-turnover}) with CadA as the enzyme, $v_{max} = (\alpha\times k_l^+)/\eta$ as the effective maximal turnover rate and $K = K_{Bl}/\eta$ the effective Michaelis constant, where $\eta = 1+\frac{k_l^+}{k_c^-} + \frac{k_l^+ k_A^-K_{Bc'}}{k_c^- k_A^+ [l']}$.

\paragraph*{Parameter estimation.}

The parameters of our quantitative model were estimated by using a trust-region reflective Newton method (MATLAB, The MathWorks, Inc.) to 
minimize the total $\chi^2$, defined by
\begin{equation}\label{EQNresidual_wt}
\chi^2(\vec{\theta}) = \chi^2_{kin}(\vec{\theta})  + \chi^2_{pH}(\vec{\theta}) + \chi^2_{lys}(\vec{\theta}) + \chi^2_{cad}(\vec{\theta})\,,
\end{equation}
with respect to the parameter vector $\vec{\theta} = (\theta_1,\dots,\theta_{M})$, where $M=14$ is the total number of model parameters, cf. Table~\ref{TABparameters}. The contribution of the kinetic data is calculated from
\begin{equation}
\chi^2_{kin}(\vec{\theta}) = \sum_{i=1}^{3} \frac{1}{N_i \overline{\Delta t_i}}\sum_{j=1}^{N_i} \Delta t_{ij} \left( \frac{ y_{ij} - \tilde{y}_{ij}(\vec{\theta}) }{\sigma_{ij}}\right)^2,
\end{equation}
where $y_{1j}$, $y_{2j}$ and $y_{3j}$ are the experimental data of the {\it cadBA} mRNA, the CadA activity and cadaverine concentration at time $t_j$, respectively. Similarly, the $\tilde{y}_{ij}(\vec{\theta})$ denote the corresponding values of the quantitative model for a given parameter set $\vec{\theta}$, and the $\sigma_{ij}$ are the standard errors of each measurement (estimates from our experiments: $\sigma_{1j}=5 $, $\sigma_{2j}=0.1$\,U/(mg protein), and $\sigma_{3j}=0.5$ mM $\forall j$). The $N_i$ are the number of datapoints of a given timeseries, $\Delta t_{ij} \equiv (t_{i(j+1)}-t_{i(j-1)})/2$ is the time between subsequent data points and $\overline{\Delta t_i}$ is the mean time between the data points in dataset $i$. The contribution of the dose-response curves are similarly defined and are exemplarily shown for the pH-dependent response
\begin{equation}\label{EQNchisqr_pH}
\chi^2_{pH}(\vec{\theta}) = \frac{1}{N_{pH}} \sum_{i=1}^{N_{pH}} \left( \frac{ A_\tau(pH_i) - \tilde{A_\tau}(pH_i,\vec{\theta}) }{\sigma_i}\right)^2\,,
\end{equation}
where $A_\tau(pH_i)$ and  $\tilde{A_\tau}(pH_i,\vec{\theta})$ are the experimental and theoretical  CadA activities after time $\tau$ at a given pH level $pH_i$ ($i=1,\dots N_{pH}$), respectively. The $\sigma_i$ denote the standard errors of the measurement and $N_{pH}$ is the total number of datapoints. The other contributions  $\chi^2_{lys}$ and  $\chi^2_{cad}$ are defined equivalently to Eq.~(\ref{EQNchisqr_pH}). However, they differ in the time at which the CadA activity was determined experimentally, i.e., $\tau_{pH} = 1.5$\,h,  $\tau_{lys} = 8$\,h, and  $\tau_{cad} = 3$\,h.

To account for the presence of local optima, and to quantify the uncertainty in the estimated parameters, we performed 1000 independent fits with randomly chosen initial parameter sets (within their physiological ranges). In Fig.~\ref{FIGresidual_parameter_corr} the final $\chi^2$ values are plotted against the final parameters. We followed Ref.~\cite{Wall_Arxiv_09} to compute the asymmetric errors $\sigma_+$ and $\sigma_-$ with respect to the optimal parameter values $\vec{\theta^{opt}}$ listed in Table~\ref{TABparameters}. The squared errors for parameter $\theta_k$ were calculated using the equations
\begin{eqnarray}\label{EQNerrors}
\sigma_{k,+}^2 & = & \frac{\displaystyle \sum_{i:\theta_{k,i}>\theta_k^{opt}} (\theta_{k,i} - \theta_k^{opt})^2 e^{-\chi_i^2/2}}{ \displaystyle\sum_{i:\theta_{k,i}>\theta_k^{opt}} e^{-\chi_i^2/2}}\,,\,\mathrm{and} \nonumber \\
\sigma_{k,-}^2 & = & \frac{\displaystyle \sum_{i:\theta_{k,i}<\theta_k^{opt}} (\theta_{k,i} - \theta_k^{opt})^2 e^{-\chi_i^2/2}}{ \displaystyle\sum_{i:\theta_{k,i}<\theta_k^{opt}} e^{-\chi_i^2/2}}\,,
\end{eqnarray}
where $\theta_{k,i}$ is the value of parameter $\theta_k$ in the $i_{th}$ fit, $\theta_k^{opt}$ is the value of $\theta_k$ in the fit with the lowest value of $\chi^2$, and $\chi_i^2$ is the value of $\chi^2$ for the $i_{th}$ fit. In using the likelihood function $e^{-\chi^2/2}$, we assume that the errors in the measurements are independent and normally distributed with widths equal to the standard error of the mean.

\section*{Acknowledgements}

It is a pleasure to thank T. Hwa and M. Saier for helpful discussions. This work was supported by the Excellence Cluster ``Nanosystems Initiative Munich'' and by the Deutsche Forschungsgemeinschaft through grants JU270/5-1 and Exc114/1. 

\bibliographystyle{unsrt}

\end{document}